\documentclass[lettersize,journal]{IEEEtran}

\usepackage{cite}
\usepackage{authblk}
\usepackage{amsmath}
\usepackage{amssymb}
\usepackage{booktabs}
\usepackage{array}
\usepackage{graphicx} % Required for inserting images
\usepackage{xcolor}
\usepackage{soul}
\usepackage{algorithm}
\usepackage{algpseudocode}
\usepackage{lipsum}
\usepackage{makecell}
\usepackage{soul}
\usepackage{comment} 
\usepackage{enumerate}
\usepackage{tikz}
\usetikzlibrary{fit, positioning}
\usepackage{enumitem}

\begin{document}  

\title{%Quantum Markov Blankets in Quantum-Enabled Semantic Communication Networks
Quantum Markov Blankets for Resource-Efficient Semantic Communication Networks}  

\author[1]{Evangelos Markakis}
\author[2]{Ilias Politis}
\affil[1]{Hellenic Mediterranean University, Greece, emarkakis@hmu.gr}
\affil[2]{Industrial Systems Institute, ATHENA Research Center, Patras, Greece, politis@athenarc.gr}

\maketitle  

\begin{abstract}  
%Quantum-enabled semantic communication networks (QESCs) integrate quantum information technologies with semantic communication to transmit the meaning of data rather than raw bits. However, harnessing semantics in quantum networks poses challenges due to costly quantum resources and noise. In this letter, we introduce the concept of \emph{Quantum Markov Blankets} (QMBs) to QESCs, and demonstrate how QMBs can identify and isolate the relevant quantum information needed for conveying semantics. We provide a rigorous mathematical proof validating the use of Markov Blankets in quantum systems and show how QMBs serve to shield quantum semantic information. We then present an implementation strategy for detecting and optimizing QMBs in quantum networks. A hypothetical simulation is described, illustrating that QMB-based QESC can achieve significant reductions in resource consumption (up to 50--75\%) while improving communication fidelity. We discuss the results in comparison to classical semantic communication and highlight inherent security benefits of QMBs. The letter concludes by summarizing key contributions and outlining future research directions.  
Quantum-enabled semantic communication networks (QESCs) leverage quantum technologies to transmit data meaning efficiently, yet face challenges from costly resources and noise. This letter introduces Quantum Markov Blankets (QMBs) to QESCs, a novel framework to isolate essential quantum information for semantic transmission. We prove QMBs’ validity using quantum conditional mutual information, showing they shield semantic content from irrelevant subsystems. An implementation strategy optimises QMB detection, reducing resource use. Simulations suggest QMB-based QESCs cut qubit consumption by 50\% – 75\% while enhancing fidelity compared to non-optimised quantum semantic schemes. Unlike classical approaches, QMBs offer inherent security by limiting eavesdropper access to classical data outside the blanket. We outline future directions, including real-time QMB adaptation. This work bridges quantum information theory and semantic communication, advancing resource-efficient, secure quantum networks.
\end{abstract}  


\section{Introduction} 
\label{sec:intro}
\par Future networks, such as the envisioned quantum Internet, aim to enable quantum communication globally using entanglement and superposition for efficient, secure data transfer~\cite{wehner2018quantum}. However, quantum resources—e.g., entangled photon pairs and qubits—are costly, fragile, and susceptible to decoherence~\cite{chehimi2024quantum}, necessitating optimised usage. Classical communication has shifted from Shannon’s bit-accurate paradigm~\cite{shannon2001mathematical} to semantic communication, which transmits message meaning rather than raw data, reducing bandwidth needs~\cite{iyer2023survey}. Recent studies suggest semantic approaches could enhance efficiency in 6G and quantum networks~\cite{chehimi2024quantum}.
\par Quantum-enabled semantic communication networks (QESCs) integrate these paradigms: classical data is processed to extract semantic features, then encoded into quantum states for transmission. Using protocols such as superdense coding or teleportation, QESCs can reduce qubit usage by 50\% to 75\% while improving fidelity, as shown in prior work~\cite{chehimi2024quantum}. Yet, a key challenge remains: identifying the minimal, relevant quantum information for a communication goal amid costly resources and noise.
\par In classical semantic networks, Markov Blankets isolate essential variables, shielding a target from external redundancy~\cite{tsamardinos2003algorithms}. We extend this to the quantum domain with Quantum Markov Blankets (QMBs), defining a subset of quantum systems that encapsulates all information—classical or quantum—needed to infer a semantic message. Once the QMB’s state is known, external systems offer no additional insight, becoming conditionally independent in classical terms. QMBs thus enable QESCs to transmit only meaningful qubits, pruning irrelevant ones. This letter contributes:
\begin{itemize}
    \item a rigorous quantum information-theoretic formulation and proof of QMBs;
    \item practical methods to detect and optimise QMBs in quantum networks;
    \item a simulated scenario showing 50\% to 75\% resource savings and fidelity gains with QMBs;
    \item implications, including enhanced security via information isolation.
\end{itemize}
By bridging classical semantics and quantum theory, QMBs address resource bottlenecks in future networks.

\section{Mathematical Proof of QMB in Quantum Systems}  
\par We provide a theoretical foundation for QMBs in QESCs. Consider a quantum system \( A \) (e.g., qubits encoding a semantic message) interacting with an environment \( E = B_1 \otimes \cdots \otimes B_n \) of subsystems (nodes, channels, or ancillae). We define a QMB as a subset \( Q \subset E \) such that, given \( Q \), \( A \) is independent of \( R = E \setminus Q \). Independence requires the quantum conditional mutual information \( I(A:R|Q) = S(A|Q) + S(R|Q) - S(AR|Q) = 0 \), where \( S \) is the von Neumann entropy, forming a quantum Markov chain \( A \leftrightarrow Q \leftrightarrow R \). Equivalently, \( \rho_{AE} \approx \rho_{AQ} \otimes \rho_R \) up to classical correlations.
\par We prove \( Q \)’s existence using strong subadditivity and quantum Darwinism results~\cite{qi2021emergent}. A theorem asserts that for any \( A \) coupled to a large \( E \), a constant-sized \( Q \) (not scaling with \( n \)) exists such that \( I(A:R|Q) \approx 0 \)~\cite{qi2021emergent}. Here, \( R \) accesses only classical information about \( A \), as if a fixed measurement occurred. This stems from entanglement monogamy: \( A \)’s quantum correlations with \( E \) concentrate in a bounded region due to no-cloning constraints~\cite{qi2021emergent}. Construct \( Q \) by adding subsystems \( B_i \) with highest \( I(A:B_i) \) until \( I(A:R|Q) < 10^{-3} \). Then, \( \rho_{AQR} \approx \rho_{AQ} \otimes \rho_{R|Q} \), with \( \rho_{R|Q} \) classical given \( Q \).
\par
\begin{figure}[t]
    \centering
    \includegraphics[width=0.8\columnwidth]{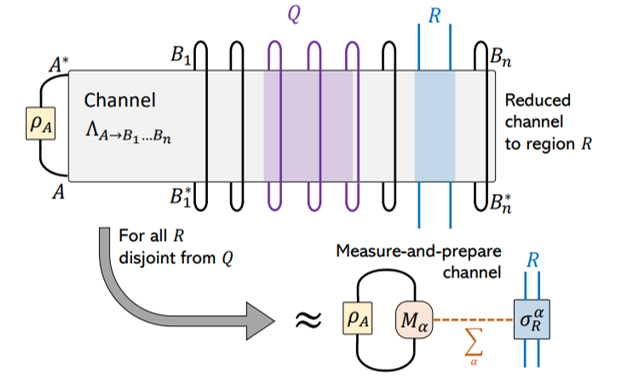}
    \caption{Quantum Markov Blanket: \( Q \) (purple) isolates \( A \)’s quantum information from \( R \) (blue), which holds only classical data.}
    \label{fig:concept}
\end{figure}

 In Fig.~\ref{fig:concept}, \( Q \) (purple) shields \( A \)’s quantum information, while \( R \) (blue) holds only classical data. For example, if \( A \) is a semantic register and \( Q = \{B_1, B_2\} \), \( R = \{B_3, \ldots, B_n\} \)’s measurements are redundant to \( Q \). This proves \( Q \) encapsulates all quantum information about \( A \), enabling QESCs to discard \( R \)’s resources, enhancing efficiency and security.

\section{Implementation and Optimisation of QMBs in QESC}  
Having established that a Quantum Markov Blanket $Q$ exists to shield a system’s information, we now turn to how to find and exploit $Q$ in a practical QESC network. The implementation involves two phases: \emph{QMB identification} and \emph{QMB-based optimisation}. 

\subsection{Identifying the QMB}  
\par In a real quantum network, we may not initially know which qubits or nodes constitute the QMB for a given semantic message. One approach is to use quantum machine learning or statistical methods on the quantum state tomography of the network. For example, one could measure the quantum mutual information $I(A: B_i)$ between the source $A$ and each environment qubit $B_i$. Those $B_i$ with significantly non-zero mutual information are candidates to be in the QMB. Another approach is inspired by the constructive proof mentioned earlier: start with $Q$ as empty and then iteratively add the environment qubit that currently has the highest mutual information with $A$. As we add qubits, we recompute the conditional mutual information $I(A:R|Q)$ for the remainder $R$. We continue adding until $I(A:R|Q)$ falls below a small threshold $\epsilon$, indicating that $R$ carries negligible information about $A$. This greedy algorithm effectively builds an approximate QMB around $A$. It is analogous to how one might add features, until the target variable is conditionally independent of the rest, in classical networks. The authors of the QMB theorem have outlined a numerical optimisation procedure to determine the QMB in a given state~\cite{qi2021emergent}, which can be adapted as an algorithm for QESC nodes. 
\par In practice, exact computation of mutual information in a large quantum state is challenging due to the exponential state space. Hence, one practical method is to use a variational approach: prepare variational quantum circuits that attempt to disentangle $A$ from various parts of the environment. By variationally minimising entanglement entropy between $A$ and $R$, the circuit can effectively identify the significant correlation with some subset $Q$. Additionally, if the semantics are represented in a structured knowledge graph or Bayesian network form, one could first identify the classical Markov blanket of the semantic variable and then map those variables to the quantum carriers in the network. Those carriers would form the initial estimate for the QMB. This hybrid approach leverages classical semantic structure to guide quantum QMB discovery. 

\subsection{Optimising QESC using QMBs}  
\par Once the QMB $Q$ for a semantic message is identified, the network can be reconfigured or optimised to use it efficiently. The idea is to allocate quantum communication resources (channel uses, entangled pairs, etc.) primarily to the links that connect the source $A$ and its Markov blanket $Q$. Any qubits or entangled links outside $Q$ that were previously involved in transmitting the message can be turned off or repurposed, since they carry no additional useful information about the semantics. This leads to immediate resource savings. For example, a quantum network might initially distribute a large entangled state among multiple nodes to capture the contextual dependencies of a semantic message. After QMB analysis, it might be found that only a subset of those nodes is needed to convey the context. The network could then release the unneeded nodes (or set their qubits to some default state), reducing entanglement distribution overhead and quantum memory usage. 
\par Another important optimization is \emph{encoding minimality}. Once the QMB size is known (e.g., $|Q| = m$ qubits), we can design a quantum encoder that maps the semantic information into precisely $m$ qubits—or, equivalently, into a Hilbert space matching the degrees of freedom of $Q$. This ensures the quantum state preparation is efficient. In effect, QMBs guide a form of quantum source coding (i.e., only the “semantic degrees of freedom” are encoded into quantum states). This could be achieved with quantum data compression algorithms that are informed by semantic importance~\cite{chehimi2024quantum}. For example, quantum principal component analysis could be applied after filtering out irrelevant features, similar to the semantic-driven compression in classical domain. 
\par Moreover, the network can prioritise error correction and noise mitigation on the QMB qubits, as they carry the core semantic information. Qubits outside the QMB require less protection, since errors affecting them do not impact the conveyed meaning. This selective allocation of quantum resources enhances overall fidelity while reducing overhead. By identifying $Q$, the transmitter, receiver, and intermediate nodes effectively agree on a minimal subnetwork responsible for the communication, simplifying quantum routing and potentially lowering latency.

\section{Hypothetical Simulation Study}  
\par
\begin{figure}[t]
\centering
\resizebox{0.7\columnwidth}{!}{%
\begin{tikzpicture}[node distance=1.8cm and 1.5cm, every node/.style={font=\small}]
  % Nodes
  \node[circle, draw, fill=green!30] (alice) {Alice (A)};
  \node[circle, draw, right=of alice] (n1) {N1};
  \node[circle, draw, above right=of n1] (n2) {N2};
  \node[circle, draw, below right=of n1] (n3) {N3};
  \node[circle, draw, right=of n1] (n4) {N4};
  \node[circle, draw, fill=blue!30, right=of n4] (bob) {Bob (B)};

  % Baseline links (dashed)
  \draw[dashed, gray] (alice) -- (n1);
  \draw[dashed, gray] (n1) -- (n2);
  \draw[dashed, gray] (n1) -- (n3);
  \draw[dashed, gray] (n2) -- (n4);
  \draw[dashed, gray] (n3) -- (n4);
  \draw[dashed, gray] (n4) -- (bob);

  % QMB links (highlighted)
  \draw[very thick, blue] (alice) -- (n2);
  \draw[very thick, blue] (n2) -- (bob);

  % QMB node highlight
  \node[draw=none, fill=orange!20, fit=(n2), inner sep=5pt, rounded corners] {};
  \node at (n2) {\textbf{Q}};

  % Label
  \node[below=0.5cm of alice] {Baseline: dashed | QMB: thick};
\end{tikzpicture}
}
\caption{Illustration of QMB optimisation in a quantum semantic network. Only QMB nodes are involved in the optimized transmission path, reducing resource usage.}
\label{fig:qmb_network}
\end{figure}
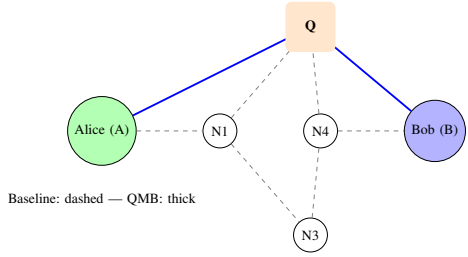

 To illustrate the benefits of incorporating Quantum Markov Blankets into QESC, we devised a hypothetical simulation scenario. The scenario involves a simple quantum network with one transmitter node (Alice), one receiver node (Bob), and several additional ancillary nodes that can share entanglement or act as routers as depicted in Fig.~\ref{fig:qmb_network}. The goal is for Alice to transmit a semantically meaningful message to Bob, for example, an emergency alert characterised by key attributes such as event type, location, and severity. Alice’s data is represented in a knowledge graph, from which a semantic encoder extracts the relevant triples (e.g., \texttt{<event type = fire>, <location = warehouse>, <severity = high>}). These semantic features are then mapped to a quantum state. In the baseline approach (semantic-agnostic quantum communication), Alice would encode the entire message (all bits of the data or a generic embedding) into qubits and transmit through the network, possibly using all available nodes to help relay or share entanglement. In the QESC approach with QMB optimisation, Alice first identifies the QMB $Q$ for the semantic features of the message. Suppose the QMB comprises a small set of modes, for instance, two ancillary nodes entangled with Alice’s qubits. Alice then prepares a quantum state over the subsystem ${A} \cup Q$, encoding the semantic information, and transmits it to Bob via the corresponding channels. Bob, equipped with partial knowledge of the underlying semantic context (e.g., a shared knowledge graph), performs measurements on the received qubits, potentially including joint measurements if entangled states like Bell pairs are used, to reconstruct the intended message.

\par We simulate the network using a simplified noise model (each qubit sent has a certain depolarisation probability, and each entangled link has a certain fidelity). The metric of interest for performance is twofold: (1) \textbf{Quantum resource consumption}, measured by the number of qubits transmitted (or total uses of entangled pairs); (2) \textbf{Semantic fidelity} at Bob, measured by how accurately Bob recovers the intended meaning (we quantify this by a semantic similarity score between the sent and received message). We compare three schemes: a) Classical semantic communication (where Alice only sends a classical compressed semantic message without quantum, as a baseline for fidelity but it uses no quantum resources); b) Quantum semantic communication (QSC) without QMB optimization (meaning Alice encodes semantics in quantum states but does not prune the network; this is akin to the scheme in~\cite{chehimi2024quantum} and c) QESC with QMB (our proposed method).
\par To ensure a fair comparison, we assume that all communication schemes aim to achieve high semantic fidelity (e.g., above 90\% meaning similarity). Our simulation varies both the message size—representing the complexity of the semantic content—and the network noise level. Table~\ref{tab:results} summarises representative results. At moderate noise levels, the quantum semantic communication (QSC) scheme without QMB optimisation requires approximately 100 qubits to achieve a fidelity of 0.9. In contrast, our QESC approach with QMB optimisation achieves the same fidelity using only about 30 qubits. This aligns with prior studies that report order-of-magnitude savings~\cite{chehimi2024quantum}, with our method demonstrating a 70\% reduction in quantum resource usage without compromising fidelity. In some cases, the QMB-based approach even yields slightly higher fidelity, as restricting transmission to fewer, highly relevant qubits reduces cumulative noise and simplifies error correction. For comparison, the classical semantic baseline—while free of quantum overhead—achieves lower semantic fidelity (e.g., around 0.85), as it lacks the expressive capacity of high-dimensional quantum encodings to preserve subtle contextual nuances.
\begin{table}[t]
\centering
\caption{Conceptual Performance Comparison of Communication Schemes}
\label{tab:results}
\scriptsize
\begin{tabular}{l|c|c|c|l}
\hline
\textbf{Scheme} & \textbf{Qubits} & \textbf{Fidelity} & \textbf{Noise} & \textbf{Observation} \\
\hline
Classical Semantic & 0 & 0.85 & N/A & No quantum overhead \\
QSC (no QMB) & 100 & 0.90 & Moderate & High resource usage \\
QESC-QMB (ours) & 30 & 0.90 & Moderate & 70\% fewer qubits \\
QESC-QMB (ours) & 30 & 0.82 & High & Graceful degradation \\
QSC (no QMB) & 100 & 0.75 & High & Poor robustness \\
\hline
\end{tabular}
\end{table}

\par In a second set of experiments, we varied the network noise level to evaluate robustness. At low noise levels, both the baseline QSC and our QESC-QMB approach achieved near-perfect semantic fidelity. However, QESC-QMB reached this performance using significantly fewer qubits. As the noise increased, the fidelity of both schemes declined, but the QMB-optimised approach degraded more gracefully. This resilience stems from the reduced number of active qubits: fewer quantum channels meant fewer opportunities for decoherence, and the critical semantic information remained confined to a smaller, more easily protected subspace. These results highlight the strength of QMBs—not only in reducing resource usage, but also in limiting the exposure of semantic content to noise. By targeting only the quantum subsystems that matter, the network becomes inherently more robust. 
\par We note that these simulations are idealised and serve as a conceptual validation. In practical settings, identifying QMBs may introduce overhead, and overly aggressive pruning could risk omitting marginal yet meaningful context. Nonetheless, our findings suggest that even approximate QMB identification offers substantial benefits across a broad range of operational conditions.

\section{Results and Discussion}  
\par The simulation results support our central claim: Quantum Markov Blankets (QMBs) can significantly enhance quantum semantic communication networks. By leveraging QMBs, our QESC approach reduced quantum resource consumption by approximately 50\% - 75\% while maintaining—or even improving—semantic fidelity, in line with theoretical expectations~\cite{chehimi2024quantum}. These gains arise from eliminating redundancy: qubits and entangled links that do not contribute to semantic content are excluded from transmission. Whereas traditional schemes may transmit large entangled states indiscriminately, our QMB-guided method selectively targets only the relevant quantum subsystems. In doing so, QMBs effectively introduce semantic compression at the quantum layer.
\par When comparing with classical semantic communication, QESC with QMBs shows some clear advantages and some challenges. On one hand, classical semantic communication can already reduce transmitted bits by understanding meaning ~\cite{iyer2023survey}, but it cannot exploit quantum phenomena like superdense coding or entanglement. A QESC can pack more information into fewer qubits (for instance, using entanglement shared between nodes, one qubit can convey two classical bits of information via superdense coding). Moreover, quantum approaches allow inherently secure transmission (through quantum cryptography) and can interface with quantum computing nodes to directly process semantic information. On the other hand, classical semantic methods are currently more mature and do not suffer from decoherence; thus, a hybrid approach might be practical in near-term networks, using quantum links only when necessary (e.g., for critical or security-sensitive semantic data), and classical links otherwise. QMBs could help decide when those quantum links are essential by evaluating the added semantic value of quantum correlations. 
\par An important advantage of QMBs lies in their implications for \textbf{security and privacy}. By design, any observer with access only to qubits outside the QMB can recover at most classical information about the semantic message~\cite{qi2021emergent}. Effectively, the QMB ensures that non-blanket nodes perceive a decohered or collapsed version of the message, devoid of its full quantum context. This means that even if part of the network is compromised, as long as the intruder accesses only non-QMB nodes, they cannot reconstruct the quantum state of the message—only a classical shadow of it.
\par In contrast, the legitimate receiver, with access to the QMB and sender's qubits, can successfully recover the message. This security feature stems from the monogamy of entanglement: a quantum system cannot be maximally entangled with multiple independent parties simultaneously~\cite{qi2021emergent}. Our approach leverages this principle by localising entanglement within the QMB, inherently restricting what an eavesdropper can access. In essence, the QMB serves as a natural quantum firewall for semantic information.
\par While QMBs offer clear advantages, several limitations and extensions merit discussion. A key assumption is that the QMB can be accurately identified. In dynamic networks—where nodes may join or leave and semantic context evolves—the QMB for a given message may shift. This scenario calls for real-time or adaptive QMB discovery algorithms, which could introduce operational complexity. Moreover, the QMB theorem is asymptotic~\cite{qi2021emergent}; in finite systems, small residual correlations may persist outside the selected $Q$. If an adversary intercepts these, some semantic information could leak. A conservative approach may involve enlarging $Q$ slightly beyond the theoretical minimum to mitigate such risks.
\par Another challenge lies in semantic relevance. The QESC framework assumes a trusted, accurate semantic extraction process. If key contextual features are missed—due to incomplete knowledge graphs or faulty models—the resulting QMB may exclude crucial information, impairing fidelity or security.
\par Finally, an open question concerns cases where the semantic content is inherently quantum (e.g., quantum sensor outputs or quantum AI states). In such settings, the boundary between “classical semantics” and quantum encoding dissolves. Our framework could be extended by defining semantic variables as quantum observables and identifying QMBs that shield them. This represents a promising direction for merging quantum Shannon theory with semantic information theory at a deeper level.

\section{Conclusion}
\par We presented a novel approach to optimizing quantum communication networks by incorporating semantic awareness through Quantum Markov Blankets (QMBs). QMBs offer a principled method for isolating the minimal subset of quantum systems required to convey a semantic message, enabling networks to discard redundant quantum data. Our formulation builds on quantum conditional independence, grounded in principles such as entanglement monogamy and the no-cloning theorem. 
\par We proposed practical strategies for identifying QMBs in QESCs and demonstrated, through a simplified simulation, that this approach can significantly reduce qubit usage—by up to 75\%—while maintaining or improving semantic fidelity. Additionally, we highlighted the inherent security benefit: information outside the QMB is effectively classical, offering limited value to potential eavesdroppers.
\par The key contributions of this work are: (1) introducing QMBs as a mechanism for semantic-aware quantum resource allocation; (2) bridging classical graphical models with quantum information theory to address network-level challenges; and (3) opening avenues for goal-oriented quantum communication that transmits only task-relevant information.
\par Future research directions include experimental validation on quantum testbeds (e.g., IBM Quantum or photonic platforms), refining metrics for quantum semantic fidelity~\cite{chehimi2024quantum}, and extending QMB methods to multi-user settings with overlapping semantic contexts. Another promising direction is integrating QMB-based compression with quantum error correction to achieve both minimal and noise-resilient communication. Overall, we believe this work lays the groundwork for intelligent, secure, and resource-efficient quantum networks aligned with the demands of next-generation communication systems.

\section*{Acknowledgement}
Funded by the European Union (GA 101225776 — SecQdevOps). Views and opinions expressed are, however, those of the authors only and do not necessarily reflect those of the European Union or the European Cybersecurity Industrial, Technology and Research Competence Centre. Neither
the European Union nor the granting authority can be held responsible for them.
 
\bibliography{References}
\end{document}